\documentstyle[12pt]{article}

\newcommand{\Dbar}{\not{\!{\!D}}}
\newcommand{\pabar}{\not{\!{\!\partial}}}
\newcommand{\Sbar}{\not{\!{\!S}}}

\newcommand{\Tr}{\mbox{Tr}}

\newcommand{\Od}{{\cal O}}

\input epsf
\begin{document}
\title{Primordial torsion fields as an explanation of the
anisotropy in cosmological electromagnetic propagation}
\author{Antonio Dobado \\
Departamento de F\'{\i}sica Te\'orica \\
Universidad Complutense de Madrid\\
 28040 Madrid, Spain\\ and \\
Antonio L. Maroto\\
Departamento de F\'{\i}sica Te\'orica \\
Universidad Autónoma de Madrid\\
 28049 Madrid, Spain}

\date{\today}

\maketitle
\begin{abstract}
In this note we provide a simple explanation of the recent
finding of anisotropy in electromagnetic (EM) propagation 
claimed by Nodland and Ralston. We
consider, as a possible origin of such  effect,
 the effective coupling between EM fields and some tiny background 
torsion field. The coupling  is obtained after integrating out charged 
fermions, 
it is gauge invariant and does not require the introduction of any new
 physics.    

\end{abstract}
\newpage
\baselineskip 0.83 true cm

In a recent paper Nodland and Ralston \cite{[1]} have reported on a 
systematic rotation
of the plane of polarization of electromagnetic radiation coming from 
distant radio
galaxies, even after Faraday rotation is extracted. Providing there is not 
any hidden
systematic bias in the data or in their analysis, this finding could 
indicate the
presence  of some anisotropic background $s^{\mu}$ over
large cosmological scales. The authors of \cite{[1]} claim that
this effect can be described by a term in the electromagnetic lagrangian 
given by
\begin{eqnarray}
(\Lambda_s^{-1}/4)\epsilon^{\mu\nu\alpha\beta}F_{\mu\nu}A_{\alpha}s_{\beta}
\end{eqnarray}
 where $A_{\mu}$ is the electromagnetic potential, 
$F_{\mu\nu}=\partial_{\mu}A_{\nu}-\partial_{\nu}A_{\mu}$,  
$\Lambda_s^{-1}$ is a constant of the order of $10^{-32}eV$ in natural units
 and $s_{\mu}=(0,\vec{s})$ is some phenomenological vector like parameter
with constant unit $\vec{s}$. From their analysis of $71$ distant 
galaxies $(z>0.3)$,
they found $\vec{s}=(decl,R.A.)=(0^o \pm 20^o,21 \pm 2 hrs)$. 

From the point of view of the physical origin of $s_{\mu}$, the most 
natural assumption
seems to be to consider it as a dynamical vector field. However, as it 
was discussed in 
\cite{[1]}, the  
main problem of that explanation is that the lagrangian above is not 
gauge invariant, 
unless $s_\mu$ is always constant, which is quite unnatural for a 
dynamical field, 
or it is the gradient of some scalar field $\phi$, 
$s_{\mu}=\partial_{\mu}\phi$. 
In this last case, new physics containing this field must be
introduced. 

After the publication of \cite{[1]} there have been several criticisms  to the
analysis of the data \cite{crtic} and responses to them \cite{resp}. In 
any case, 
the purpose of this paper is to point out that, provided the effect exists, 
there is a simple interpretation of the  lagrangian above
within the framework of standard quantum electrodynamics (QED) and 
classical gravity.
In this interpretation, 
$s_{\mu}$ is understood as the pseudotrace of the torsion field and
 gauge invariance is preserved without the
introduction of any new scalar field $\phi$. In order to show that, we 
consider the
following simple model to describe the present state of the observable 
universe. We
assume a flat space-time with a classical background torsion 
field $T^{\mu\nu\alpha}$
(although our results can be generalized to the case when curvature is
present in a straightforward way).
The mathematical origin of this tensor is the following: consider a 
pseudo-Riemannian 
space-time manifold with metric tensor $g_{\mu\nu}$. As usual, in order to
define the parallel transport of vectors, we should introduce a new object, an
affine connection, whose components are 
$\hat \Gamma^{\lambda}_{\;\;\mu\nu}$. Such 
an arbitrary connection
is  in principle independent of the metric. However if we want the
lengths and angles of vectors to be invariant under parallel transport, it is
needed that the connection is metric, that is:
\begin{eqnarray}
(\hat\nabla_\lambda g)_{\mu\nu}=
\partial_\lambda g_{\mu\nu}-\hat\Gamma^{\kappa}_{\lambda\mu}g_{\kappa\nu}-
\hat\Gamma^{\kappa}_{\lambda\nu}g_{\kappa\mu}=0
\end{eqnarray}
where $\hat\nabla$ is the corresponding covariant derivative. 
This condition allows us to find the following general form for this kind 
of connections:
\begin{eqnarray}
\hat \Gamma^{\lambda}_{\;\;\mu\nu}=\Gamma^{\lambda}_{\;\;\mu\nu}
+\frac{1}{2}\left(T_{\nu\;\;\mu}^{\;\lambda}+T_{\mu\;\;\nu}^{\;\lambda}-
T_{\;\;\mu\nu}^{\lambda}\right)
\label{torchr}
\end{eqnarray}
where the antisymmetric part, 
$T_{\;\;\mu\nu}^{\lambda}=\hat\Gamma_{\;\;\mu\nu}^{\lambda}
-\hat\Gamma_{\;\;\nu\mu}^{\lambda}$
is known as the torsion tensor and $\Gamma^{\lambda}_{\;\;\mu\nu}$
are the usual Christoffel symbols that can be obtained from the metric tensor. 
In absence of torsion, any metric connection reduces to the 
Levi-Civita one, given by
the Christoffel symbols. This was the connection considered by Einstein in his
formulation of General Relativity. However nowadays the modern theories of
gravity consider the metric tensor and the connection as independent 
entities and 
therefore torsion appears in a natural way.

By means of the Einstein Equivalence Principle, it
is possible to find the minimal lagrangian for a fermion interacting 
with a background
gravitational field with torsion \cite{[3]}, then it is easy to show that
only the axial pseudo-trace  vector-field
$S_{\beta}=\epsilon_{\mu\nu\alpha\beta}T^{\mu\nu\alpha}$ appears in 
the minimal coupling.  
When there is also a background
electromagnetic field and no curvature, the lagrangian  describing the 
electrodynamics 
of different charged 
fermions $\Psi_i$ is given by  
\begin{eqnarray}
{\cal
L}=-\frac{1}{4}F_{\mu\nu}F^{\mu\nu}+\sum_{i}\overline\Psi_i(i\Dbar_i - m_i
-\frac{1}{8}\gamma_5 \Sbar)\Psi_i
\end{eqnarray}
where $D^{\mu}_i=\partial^{\mu}-ieQ_iA^{\mu}$ with $m_i$ and $Q_i$ being  
the mass 
and the electric 
charge  of the corresponding fermionic field. Now it is
immediate to integrate out the fermionic fields  to find the effective 
action for the 
electromagnetic field propagating in the torsion classical background: 
\begin{eqnarray}
e^{iS_{eff}[A,S]}=\int [d\Psi][ d\overline \Psi] e^{i\int d^4x {\cal L}}
\label{effa}
\end{eqnarray}
where $\Psi$ denotes all the fermion fields. By using standard 
manipulations, it is possible
to write:
\begin{eqnarray}
S_{eff}[A,S]&=&-\int d^4x\left(\frac{1}{4}F_{\mu\nu}F^{\mu\nu}\right)
+i\sum_i \Tr\log(i\Dbar_i - m_i
-\frac{1}{8}\gamma_5 \Sbar)
\end{eqnarray}
which can be developed as:
\begin{eqnarray}
S_{eff}[A,S]=-\int d^4x\left(\frac{1}{4}F_{\mu\nu}F^{\mu\nu}\right)
+i\sum_i\sum_{k=1}^{\infty}
\frac{(-1)^k}{k}\Tr\left[(i\pabar-m_i)^{-1}\left(eQ_i
-\frac{1}{8}\Sbar\gamma_5\right)\right]^k
\nonumber \\
\end{eqnarray}
Computing the functional traces by using dimensional regularization  
with $D=4-\epsilon$, 
it is
possible to obtain:
\begin{eqnarray}
S_{eff}[A,S]&=&-\int d^4x\left(\frac{1}{4}F_{\mu\nu}F^{\mu\nu}\right)
\nonumber \\
&-&\sum_i\frac{e^2Q_i^2m_i^2}{2}
\int d\tilde q \;d\tilde p\; dx\; dy\; dz\;e^{ip(y-x)}
e^{iq(z-x)}\Gamma(3-D/2)\nonumber\\
&\times&\int_0^1dx_1\int_0^{x_1}dx_2\frac{i}{16\pi^2}\epsilon^{\mu\nu\rho\sigma}
\left(A_\nu^yA_\rho^z S_\sigma^x(p_\mu x_1+q_\mu x_2)\right.
\nonumber \\&+&\left.
A_\mu^yA_\rho^z S_\sigma^x(p_\nu( x_1-1)+q_\nu x_2)+
A_\mu^yA_\nu^z S_\sigma^x(p_\rho (x_1-1)+q_\rho(x_2-1))\right)\nonumber\\
&\times& F(p,q,x_1,x_2;m_i)
+\Od(S^2)+\Od(A^3)
\label{ae}
\end{eqnarray}
with
\begin{eqnarray}
F(p,q,x_1,x_2;m_i)=\left(-m_i^2+p^2(x_1-x_1^2)
+2pq(x_2-2x_1x_2)+q^2(x_2-x_2^2)\right)^{-1}
\end{eqnarray}
where $d\tilde p=d^D p \mu^{\epsilon}/{(2\pi)^D}$ and $\Od(S^2)$, 
$\Od(A^3)$ denote terms with two 
or more torsion fields and 
three or more photon fields respectively. As can be easily seen, 
this contribution to the
effective action is finite. Therefore it is possible to 
expand the function $F$ in powers of ${\cal K}^2/m_i^2$, 
where ${\cal K}$ denotes generically the
external momenta. To lowest order we find:
\begin{eqnarray}
F(p,q,x_1,x_2;m_i)=-\frac{1}{m_i^2}\left(1
+\Od\left(\frac{{\cal K}^2}{m_i^2}\right)\right)
\end{eqnarray}
In this way, we can perform explicitly the integrals in the 
Feynman parameters $x_1$ and
$x_2$ in (\ref{ae}) order by order, so that we obtain a local expansion for the 
effective lagrangian, whose lowest order is given by:
\begin{eqnarray}
{\cal
L}_{eff}=-\frac{1}{4}F_{\mu\nu}F^{\mu\nu}
-\frac{\alpha}{24\pi}\left(\sum_i Q^2_i\right)
\epsilon^{\mu\nu\alpha\beta}F_{\mu\nu}A_{\alpha}S_{\beta}+hdt
\label{eff}
\end{eqnarray}
where $\alpha$ is fine structure constant and $hdt$ stands for 
the higher derivative 
terms coming 
from the higher orders in the expansion of the function $F$. At 
this point it is 
important to stress that, as far as this model 
is not anomalous \cite{[4]}, the effective action defined in 
(\ref{effa}) is gauge invariant.
The higher derivative terms in (\ref{eff}) can be neglected at long 
distances, but are needed to preserve the gauge invariance of the whole 
effective lagrangian. Note that, as discussed above, the second term 
in the r.h.s. of (\ref{eff}) alone is not gauge invariant unless $S_\mu$
were the gradient of some new scalar function \cite{[2]}. 

 Thus from the Feynman diagram in Fig.1 we have found the 
appropriate term in the lagrangian
in an invariant manner without the introduction of any new scalar field $\phi$,
as can be observed by the simple identification
\begin{eqnarray}
\Lambda_s^{-1}s_{\mu}=-\frac{\alpha }{6\pi }\left(\sum_i Q^2_i\right)S_{\mu}
\end{eqnarray}
For example, by considering the matter content of the Standard Model 
we have $\sum_i
Q^2_i=N_f(1+N_c(4/9+1/9))$ which for three families $(N_f=3)$ 
and three colors ($N_c=3$)
equals $8$. Thus by assuming an space-like $S_{\mu}$  field
($S_{\mu}=(0, \vec{S})$), the required module for $\vec{S}$ to explain the
polarization effect observed by Nodland and Ralston should be something about
$10^{-30}eV$. 

Concerning the origin of the torsion field, as commented before, in most of
theories of gravitation the affine connection and the vielbein 
(which is related with the
metric tensor), are considered as independent entities and thus 
torsion appears in a
natural way. If the description provided here for the Nodland 
and Ralston effect
were appropriate, it could be the first indication of the
presence of torsion at  cosmological scales. From the point of 
view of a quantum field theory for torsion, this probably requires 
that the $S_\mu$ field is masless or having a vacuum expectation value. This
could be in contradiction with the usual assumptions of low-energy effective 
quantum gravity in which the 
torsion field is supposed to have a mass of the order of Planck mass 
\cite{[5]}. In spite of that and provided that this new effect is
confirmed, we really believe that this is the most
economical explanation of the finding by Nodland-Ralston since 
it does not require the
introduction of new fields, apart 
from those appearing in the Standard Model and
our current description of classical gravity.

\vspace{0.5cm}
{\bf Acknowledgements}

We are grateful to B. Nodland for 
his comments to the preliminary version of this paper. 
This work has been supported in part by the Ministerio de Educaci\'on y
Ciencia (Spain) (CICYT AEN96-1634).

\newpage


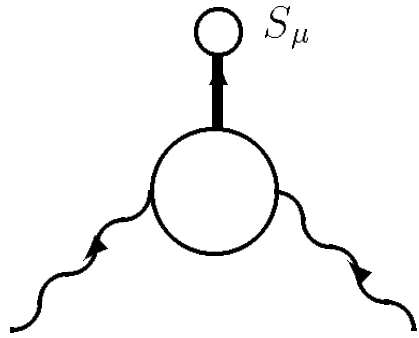
\end{document}